# Nucleation of dislocations and their dynamics in layered oxides cathode materials during battery charging

A. Singer[1], S. Hy[2], M. Zhang[2], D. Cela[1], C. Fang[2], B. Qiu[3], Y. Xia[3], Z. Liu[3], A. Ulvestad[4], N. Hua[1], J. Wingert[1], H. Liu[2], M. Sprung[5], A. V. Zozulya[5,†], E. Maxey[6], R. Harder[6], Y.S. Meng[2], O. G. Shpyrko[1]

[1]Department of Physics, University of California-San Diego, La Jolla, California 92093, United States

[2]Department of NanoEngineering, University of California-San Diego, La Jolla, California 92093, United States

[3]Ningbo Institute of Materials Technology and Engineering (NIMTE), Chinese Academy of Sciences, Zhejiang 315201, China

[4]Materials Science Division, Argonne National Laboratory, Argonne, Illinois 60439, United States

[5]Deutsches Elektronen-Synchrotron DESY, Notkestr. 85, D-22607 Hamburg, Germany

[6]X-ray Science Division, Advanced Photon Source, Argonne, Illinois 60439, United States

†Present address: European XFEL GmbH, Holzkoppel 4, 22869 Schenefeld, Germany

*Abstract:*

Defects and their interactions in crystalline solids often underpin material properties and functionality[1] as they are decisive for stability[1–5], result in enhanced diffusion[6], and act as a reservoir of vacancies[7]. Recently, lithium-rich layered oxides have emerged among the leading candidates for the next-generation energy storage cathode material, delivering 50 % excess capacity over commercially used compounds. Oxygen-redox reactions are believed to be responsible for the excess capacity[8], however, voltage fading has prevented commercialization of these new materials. Despite extensive research the understanding of the mechanisms underpinning oxygen-redox reactions and voltage fade remain incomplete. Here, using operando three-dimensional Bragg coherent diffractive imaging[2,9], we directly observe nucleation of a mobile dislocation network in nanoparticles of lithium-rich layered oxide material. Surprisingly, we find that dislocations form more readily in the lithium-rich layered oxide material as compared with a conventional layered oxide material, suggesting a link between the defects and the anomalously high capacity in lithium-rich layered oxides. The formation of a network of partial dislocations dramatically alters the local lithium environment and contributes to the voltage fade. Based on our findings we design and demonstrate a method to recover the original high voltage functionality. Our findings reveal that the voltage fade in lithium-rich layered oxides is reversible and call for new paradigms for improved design of oxygen-redox active materials.

The lithium-rich layered oxide (LRLO) compounds are among the most promising positive electrode materials for next-generation batteries. They exhibit high capacities of >300 mAhg$^{-1}$ due to the unconventional participation of the oxygen anion redox in the charge compensation mechanism[10–15], suggested by localization of 2p O electrons[14,15] or by formation of $Li_2O$[16] or of O-O peroxide dimers during operation[17]. The LRLO material is a composite of a classical layered oxide $LiTMO_2$ (with $R\bar{3}m$ space group, TM stands for Ni, Mn, Co) and $Li_2TMO_3$ (with C2/*m* space group). The local composition of cations and their associated interaction turns out to be crucial for voltage stability and lithium diffusion capabilities[12,18,19]. Various stacking sequences of the transition metal (TM) layers have been identified (stacking faults)[18,20]. X-ray and neutron scattering measurements revealed that the layer spacing (*c* lattice parameter) expands during charge from 14.25 Å to about 14.40 Å with simultaneous contraction of the *a-b* plane[20], resulting in significant volume changes and cracking of secondary particles upon the initial charge to 4.5 V[21]. Indeed, the formation of cracks and stress-induced damage in secondary particles has been identified in other layered oxides and is believed to be a leading cause of degradation and debilitating electrochemical performance[3,22,23]. The tens of microns large secondary particles are agglomerates of much smaller (sub micron) single crystalline primary particles and such morphologies are introduced to improve the volumetric energy density of the cathode (see Supplementary Fig. S1).

Despite their fundamental importance, the understanding of structural evolution during charge in the *primary* particles remains challenging. Particularly the nature of crystal defects, which are expected to occur with Burgers vectors parallel to the layers during operation[3,4], remains elusive due to the difficulty in capturing their operando formation and dynamics[2,24]. Here we directly capture the nucleation of a dislocation network in primary nanoparticles of the high capacity LRLO material $Li_{1.2}Ni_{0.133}Mn_{0.533}Co_{0.133}O_2$ during electrochemical charge (lithium extraction) (see Fig. 1). By using the *in situ* Bragg coherent diffractive imaging (BCDI) technique[2], we map the 3D displacement field inside primary battery particles during battery operation (see Supplementary Information and Figs. S2 and S3). Dislocations give rise to singularities in the displacement field, which are easily recognizable even in the presence of noise[25]. We find that while the pristine nanoparticle shows no such singularities in the present geometry (see Fig. 1), when charged to a voltage of 4.3 V versus (vs.) $Li^+$ it contains two dislocations in the bulk of the nanoparticle, revealing the formation of dislocations in bulk during charge. The dislocation density increases upon subsequent charge, and a dislocation network emerges at 4.4 V vs. $Li^+$. We identify the crystal imperfections as line defects with the Burgers vector having a component along the *c*-axis direction, perpendicular to the layers, resembling Frank's partial dislocations[1].

The *in situ* BCDI technique allows us to directly image the interior of a nanoparticle during lithium extraction (battery charge) operando (see Fig. 2). In the beginning of the charge at 4.0 V, we observe a continuous displacement field. At 4.2 V the displacement field has changed only slightly, an expected structural response of the nanoparticle to delithiation[20]. When charged to 4.3 V, before the voltage plateau that signifies oxygen evolution[16], the displacement field is qualitatively different and shows two singularities characteristic of dislocations. The discontinuity in the displacement field around a singularity is the projection of the Burgers vector along the scattering vector $q_{003}$, which matches the thickness of a single TM or lithium layer (see Fig. 1 for a schematic). Further charging to 4.4 V vs. $Li^+$ induces additional dislocations, while the two edge dislocations present at 4.3 V move slightly. By analyzing the dislocation type in various particles during lithium extraction, we find the presence of edge, screw, and mixed dislocations. We are unable to perform BCDI at voltages higher than 4.4 V possibly due to growing disorder in the structure; however, further broadening of the Bragg diffraction suggests continued formation of dislocations (see Figs. S4, S5).

The strain field distribution yields further insights into the physical and electrochemical processes underpinning the formation of dislocations (see Fig. 2b). The strain inhomogeneity in the slice shown increases between 4.0 V and 4.2 V and the strain is larger towards the bottom left edge of the particle, suggesting elevated delithiation. At 4.3 V a plastic deformation occurs via nucleation of two dislocations in bulk. Around these dislocations, the strain is compressive (reduced lattice spacing **d**) on the right and tensile on the left, revealing extra half-planes inserted from the right in agreement with the displacement field analysis (see schematic in Fig. 1). Modeling of strain profiles indicates that most of the dislocations are mixed c and a-b dislocations (see Supplementary Information and Fig. S6). We have conducted measurements on two different nanoparticles, and both of them show the formation of multiple dislocations during charge (see Fig. S7).

To have a direct comparison between LRLO and currently commercialized layered oxides, we have conducted operando BCDI on a classical layered oxide material ($LiNi_{0.80}Co_{0.15}Al_{0.05}O_2$, "NCA") (see Fig. 3). Strikingly, although we find that dislocations form in the classical material as well, their number is significantly less than in LRLO (in a different NCA nanoparticle no dislocations formed during delithiation to 4.8 V). In the nanoparticle of the classical material a single dislocation forms at 4.2 V; however, no new dislocations appear above 4.2 V. We calculate the partial strain energy in the (001) direction via $E_{p,001} = \frac{1}{2} Y \int \varepsilon_{001}(x)^2 dx$, where Y is Young's modulus[26], $\varepsilon_{001}(x)$ is the measured strain shown in Figs. 2b and 3b, $x$ is the spatial coordinate, and we integrate over the particle volume. The strain energy for LRLO materials grows monotonically up to 4.4 V, while it peaks at 4.2 V for NCA and reduces upon further lithium extraction. The strain calculations shown in Fig. 4 are in excellent agreement with Williamson-Hall analysis of powder diffraction data, which shows a gradual increase in the microstrain during charge from 4.0 V to 4.5 V for LRLO and a peak at 4.3 V for NCA (see Supplementary Fig. S5). The agreement between the strain measurements in single nanoparticles in operando BCDI and bulk powder diffraction data confirms that the particles shown in Figs. 2 and 3 are representative of the ensembles.

We attribute the origin of dislocation formation in LRLO to the limited lithium ion diffusion at high voltages, which drops from $D=10^{-14}$cm$^2$/s at 4.0 V below $D=10^{-15}$cm$^2$/s at 4.4 V Ref.[27]. Using Fick's law $\langle x^2 \rangle = 4D\tau$, where $\tau$ is the diffusion time, we determine an average travel distance of lithium ions to be $\sqrt{\langle x^2 \rangle}$~40 nm in 1 hour, which is an order of magnitude smaller than the size of the nanoparticles. Lithium extraction during charge coupled with slow diffusion will result in lithium-depleted regions at the particle boundary. We argue that in LRLO the dislocations form as a result of strain concentrators due to the volume difference between regions with low and high lithium concentrations. In the classical layered oxide materials, the diffusion rate is higher and the ductility is lower[28], which results in the dramatic difference between the rates of dislocation formation in the LRLO and classical layered materials. By tracking the location of two dislocations that formed at 4.3 V in LRLO (see arrows in Fig. 2), we estimate the speed of dislocations to be on the order of 10 nm/h. This speed is comparable with lithium mobility, indicating ionic diffusion as the dominant reason for dislocation motion.

By measuring the length of dislocations in single nanoparticles, we directly calculate the dislocation density to be $1\times10^{10}$ cm$^{-2}$ in LRLO at 4.4 V (note that in NCA the density is approximately one order of magnitude smaller, see left inset in Fig. 4). At such a high value, we anticipate a sizable impact on the material's performance, particularly on voltage fade and oxygen activity. The nucleation of line defects dramatically modifies the local lithium environment, which we show is crucial in determining the voltage. The formation of partial dislocation perturbs the superstructure stacking sequence[1], and we directly confirm the decline of the superstructure peak in *in situ* x-ray diffraction starting at 4.3 V (see right inset in Fig. 4). Even if the dislocation disappears by moving out of the particle or by annihilation with other dislocations, the superstructure is unlikely to recover once a dislocation has formed. During the subsequent cycling, the superstructure gradually disappears, and the material becomes partially disordered. The highly cycled material is trapped in a metastable state with energetically unfavorable local lithium environment. Based on the mechanistic description above we design a path to re-order the superstructure by high-temperature annealing (>150 °C). The superstructure recovery (see Fig. 5**b**) is decisive in restoring the original voltage profile (see Figs. 5**a** and 5**c**).

The experimental observation of significantly higher rate of dislocation formation in LRLO as compared with classical materials resonates with the anomalous anionic activity in LRLO. The current understanding of the anionic activity in LRLO materials includes three processes, 1: reversible oxidation of $O^{2-}$ to $O^-$, 2: further partially reversible oxidation to O, and 3: irreversible release of $O_2$ gas from the bulk material to the surface[8]. Steps 2 and 3 require elevated oxygen mobility in bulk. While enhanced 'pipe' diffusion occurs along dislocations[6], recent theoretical works suggest a slower transport of oxygen along dislocations due to cation charge accumulation at the defect site[29]; oxygen mobility due to thermal fluctuations is predicted to be negligible at room temperature. However, an external electrical current leads to reversible accumulation or depletion of oxygen vacancies at dislocations in $SrTiO_3$ single crystals as well as epitaxial thin films[7]. Similarly, we posit that during charge the lithium and electron extraction activates the emergent dislocation network in LRLO for the transport of $O^-$ and O. The dislocation-mediated oxygen mobility also likely assists the re-intercalation of lithium ions into the transition metal layer[30].

**Acknowledgements**
We acknowledge Kamila Wiaderek for providing the potentiostat during the measurements at the Advanced Photon Source and Hao Liu and Karena Chapman for collecting the ex situ powder diffraction data on the NCA material. We also thank Anton Van der Ven and Max Radin for discussions. The x-ray imaging was supported by the U.S. Department of Energy (DOE), Office of Science, Office of Basic Energy Sciences, under contract DE-SC0001805 (A.S., D.C., J.W., N.H., and O.G.S.). S.H., C.F., M.Z., H.L., and Y.S.M. acknowledge support on the materials synthesis, electrochemical and materials characterization from the NorthEast Center for Chemical Energy Storage (NECCES), an Energy Frontier Research Center funded by the U.S. Department of Energy, Office of Science, Basic Energy Sciences under Award no. DE-SC0012583. This research used resources of the Advanced Photon Source, a U.S. DOE Office of Science User Facility operated for the DOE Office of Science by Argonne National Laboratory under contract no. DE-AC02-06CH11357. We thank the staff at Argonne National Laboratory and the Advanced Photon Source for their support. Parts of this research were carried out at the light source PETRA III at DESY, a member of the Helmholtz Association (HGF). The data are stored at Sector 34-ID-C of the Advanced Photon Source and at PETRA III at DESY.

**Author Contributions.**

A.S., S.H., Y.S.M., and O.G.S. conceived of the idea. A.S., S.H., D.C., C.F., A.U., J.W., conducted the imaging experiments on LRLO nanoparticles, with assistance from E.M and R.H.. A.S. and N.H. performed the imaging experiments on NCA nanoparticles, with assistance from A.Z. and M.S.. S.H., C.F., H.L. prepared the LRLO and NCA samples and performed the materials characterization and electrochemistry testing. A.S. analyzed the imaging experiments with help from D.C.. M.Z. performed the microstrain analysis with help from Y.S.M.. B.Q, Y.X and Z.L. performed the superstructure restoration and subsequent electrochemistry testing. E.M. designed the sample environment. A.S. wrote the paper with input from all authors.

**Figures:**

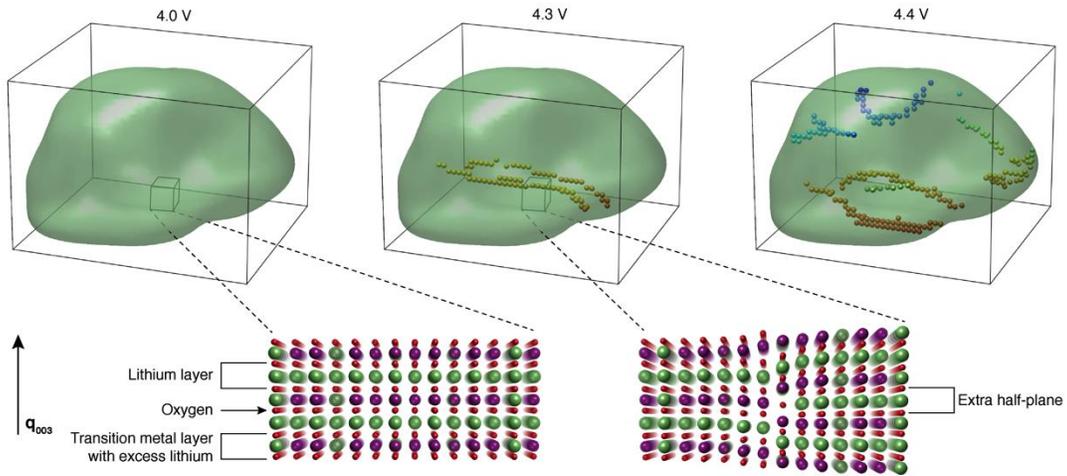

**Figure 1: Formation of a dislocation network during charge.** Isosurface representations of a LRLO particle measured in operando during charge. At a charge state of 4.0 V vs Li$^+$ no dislocations are observed in the particle. At 4.3 V two edge dislocations have formed during lithium extraction (shown by dotted lines in the particle) and at 4.4 V a dislocation network emerges (colors are used to represent different dislocations). The direction of the scattering vector $q_{003}$ is indicated and the size of the particle is around 300x300x500 nm$^3$. The dislocations have a component of the Burgers vector along the $q_{003}$ vector and a schematic of the dislocation is shown at the bottom.

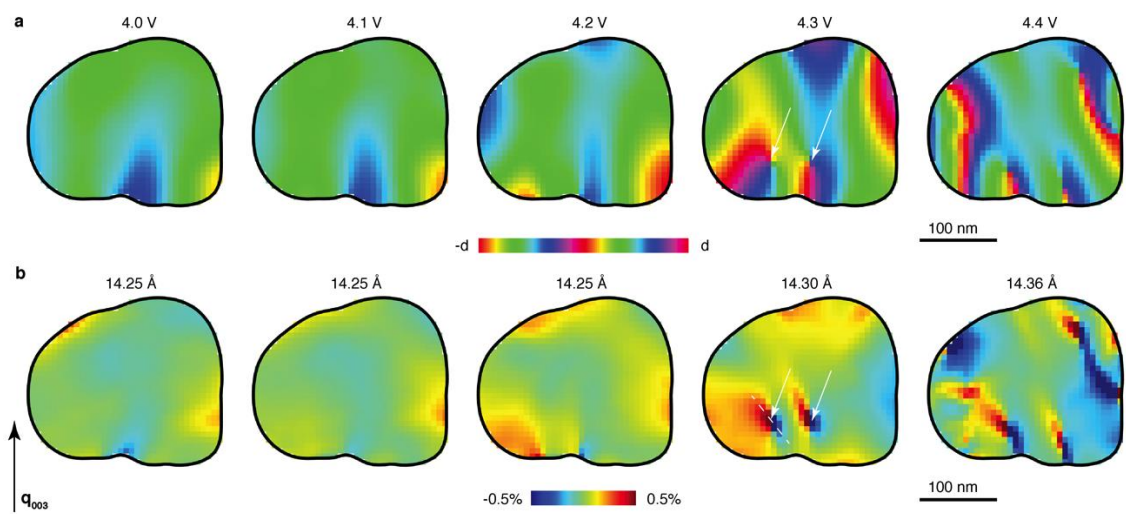

**Figure 2: In situ evolution of a LRLO nanoparticle during electrochemical charge. a** The changes in the displacement field along $q_{003}$ in a plane through the nanoparticle during charge (the plane is shown in Fig. S2). The voltage vs. Li$^+$ is indicated at the top. Two edge dislocations are identified as singularities of the displacement field and are indicated at 4.3 V vs. Li$^+$ (see schematic in Fig. 1). Additional dislocations emerge at 4.4 V vs. Li$^+$. **b** The strain along the (001) direction (perpendicular to the layers) inside of the nanoparticle calculated from the 3D displacement fields in **a**. The strain is shown around the average lattice constant indicated at the top. White arrows in **b** indicate the positions of the edge dislocations at 4.3 V vs. Li$^+$.

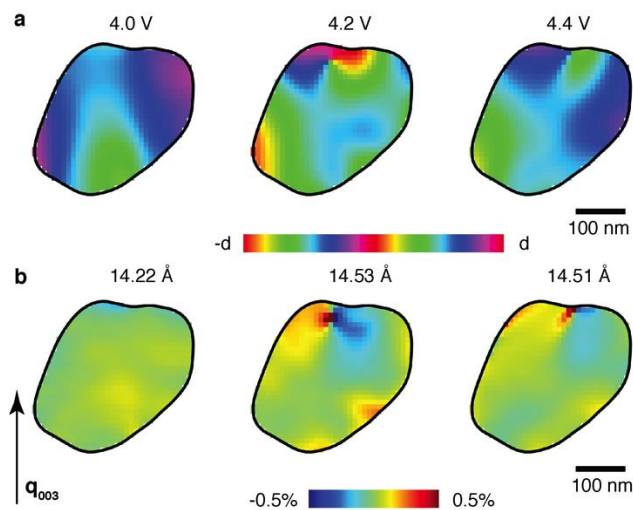

**Figure 3: In situ evolution of a NCA nanoparticle during charge.** The displacement field **a** and strain **b** along the (001) direction (perpendicular to the layers) of a single NCA particle captured *in situ* during charge. The voltages and average lattice constants are indicated.

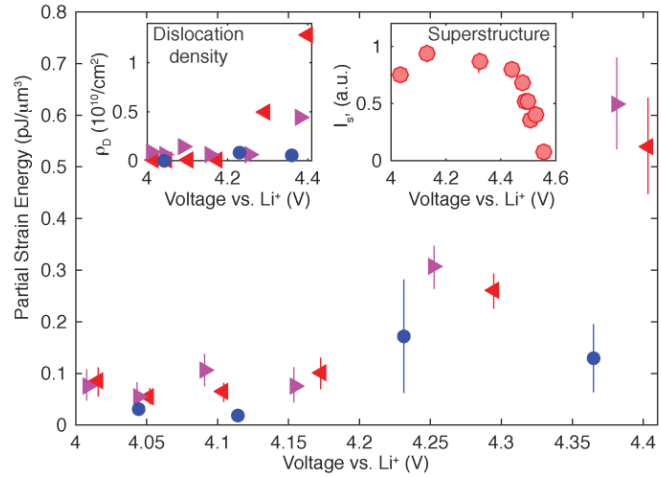

**Figure 4: Strain energy landscape of single particles of layered oxides.** Partial strain energy $E_{p,001}$ for two LRLO particles shown in Figs. 1,2 (red triangles) and Fig. S7 (magenta triangles), and a NCA particle shown in Fig. 3 (blue circles). The left inset shows the operando evolution of the dislocation density $\rho_0$ for the same particles. The right inset shows the evolution of the superstructure peak intensity during the first charge measured *in situ* using x-ray diffraction from a large number of particles.

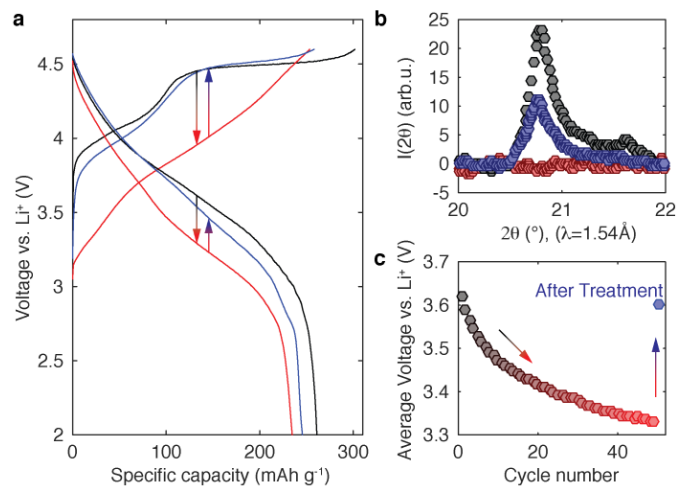

**Figure 5: A path to restore the voltage in the lithium-rich oxide material. a** Charge-discharge voltage curves of Li-rich layered cathode with Li-metal as anode for the voltage range of 2.0-4.6 V vs. Li$^+$/Li: the first cycle (black lines), 50th cycle (red lines), and the cycle after heat treatment (blue lines) are shown (the sequence of measurements is indicated by arrows). **b** The evolution of the superstructure peak intensity in the pristine state (black symbols), after 50 cycles (red symbols), and after the heat treatment (blue symbols). **c** The average discharge voltage for the first 50 cycles (black to red symbols) and after the heat treatment (blue symbol). The discharge voltage is recovered after the treatment.

**References:**
1. Hull, D. & Bacon, D. J. in *Introduction to Dislocations* iv (Elsevier, 2011). doi:10.1016/B978-0-08-096672-4.00011-6
2. Ulvestad, A. *et al.* Topological defect dynamics in operando battery nanoparticles. *Science (80-. ).* **348,** 1344–1347 (2015).
3. Wang, H., Jang, Y. I. & Huang, B. TEM Study of Electrochemical Cycling- Induced Damage and Disorder in LiCoO2 Cathodes for Rechargeable Lithium Batteries. *J. Electrochem. Soc.* **146,** 473–480 (1999).
4. Gabrisch, H., Yazami, R. & Fultz, B. The Character of Dislocations in LiCoO[sub 2]. *Electrochem. Solid-State Lett.* **5,** A111 (2002).
5. Yan, P. *et al.* Intragranular cracking as a critical barrier for high-voltage usage of layer-structured cathode for lithium-ion batteries. *Nat. Commun.* **8,** 1–9 (2017).
6. Poirer, J.-P. *Creep of crystals*. (Cambridge University Press, 2005).
7. Szot, K., Speier, W., Bihlmayer, G. & Waser, R. Switching the electrical resistance of individual dislocations in single-crystalline SrTiO3. *Nat. Mater.* **5,** 312–320 (2006).
8. Delmas, C. Battery materials: Operating through oxygen. *Nat. Chem.* **8,** 641–643 (2016).
9. Pfeifer, M. A., Williams, G. J., Vartanyants, I. A., Harder, R. & Robinson, I. K. Three-dimensional mapping of a deformation field inside a nanocrystal. *Nature* **442,** 63–66 (2006).
10. Lee, J. *et al.* Unlocking the potential of cation-disordered oxides for rechargeable lithium batteries. *Science* **343,** 519–22 (2014).
11. Sathiya, M. *et al.* Reversible anionic redox chemistry in high-capacity layered-oxide electrodes. *Nat. Mater.* **12,** 827–35 (2013).
12. Sathiya, M. *et al.* Origin of voltage decay in high-capacity layered oxide electrodes. *Nat. Mater.* **14,** 230–8 (2015).
13. Nitta, N., Wu, F., Lee, J. T. & Yushin, G. Li-ion battery materials: present and future. *Mater. Today* **18,** 252–264 (2015).
14. Seo, D.-H. *et al.* The structural and chemical origin of the oxygen redox activity in layered and cation-disordered Li-excess cathode materials. *Nat. Chem.* **8,** 692–697 (2016).
15. Luo, K. *et al.* Charge-compensation in 3d-transition-metal-oxide intercalation cathodes through the generation of localized electron holes on oxygen. *Nat. Chem.* **8,** 684–691 (2016).
16. Hy, S., Felix, F., Rick, J., Su, W.-N. & Hwang, B. J. Direct in situ observation of Li2O evolution on Li-rich high-capacity cathode material, Li[Ni(x)Li((1-2x)/3)Mn((2-x)/3)]O2 ($0 \leq x \leq 0.5$). *J. Am. Chem. Soc.* **136,** 999–1007 (2014).
17. McCalla, E. *et al.* Visualization of O-O peroxo-like dimers in high-capacity layered oxides for Li-ion batteries. *Science* **350,** 1516–21 (2015).
18. Seymour, I. D. *et al.* Characterizing Oxygen Local Environments in Paramagnetic Battery Materials via 17 O NMR and DFT Calculations. *J. Am. Chem. Soc.* jacs.6b05747 (2016). doi:10.1021/jacs.6b05747
19. Croy, J. R., Balasubramanian, M., Gallagher, K. G. & Burrell, A. K. Review of the U.S. Department of Energy's 'Deep Dive' Effort to Understand Voltage Fade in Li- and Mn-Rich Cathodes. *Acc. Chem. Res.* **48,** 2813–2821 (2015).
20. Liu, H. *et al.* Operando Lithium Dynamics in the Li-Rich Layered Oxide Cathode Material via Neutron Diffraction. *Adv. Energy Mater.* **6,** (2016).
21. Chen, C.-J. *et al.* The Origin of Capacity Fade in the Li 2 MnO 3 ·Li M O 2 ( M = Li, Ni, Co, Mn) Microsphere Positive Electrode: An Operando Neutron Diffraction and Transmission X-ray Microscopy Study. *J. Am. Chem. Soc.* **138,** 8824–8833 (2016).
22. Watanabe, S., Kinoshita, M., Hosokawa, T., Morigaki, K. & Nakura, K. Capacity fade of LiAlyNi1−x−yCoxO2 cathode for lithium-ion batteries during accelerated calendar and cycle life tests (surface analysis of LiAlyNi1−x−yCoxO2 cathode after cycle tests in restricted depth of discharge ranges). *J. Power Sources* **258,** 210–217 (2014).
23. Gent, W. E. *et al.* Persistent State-of-Charge Heterogeneity in Relaxed, Partially Charged Li 1− x Ni 1/3 Co 1/3 Mn 1/3 O 2 Secondary Particles. *Adv. Mater.* **28,** 6631–6638 (2016).
24. Huang, J. Y. *et al.* In situ observation of the electrochemical lithiation of a single $SnO_2$ nanowire electrode. *Science* **330,** 1515–20 (2010).
25. Nye, J. F. & Berry, M. V. Dislocations in Wave Trains. *Proc. R. Soc. A Math. Phys. Eng. Sci.* **336,** 165–190 (1974).
26. Qi, Y., Hector, L. G., James, C. & Kim, K. J. Lithium Concentration Dependent Elastic Properties of Battery Electrode Materials from First Principles Calculations. *J. Electrochem. Soc.* **161,** F3010–F3018 (2014).
27. Yu, H. *et al.* Electrochemical kinetics of the 0.5Li2MnO3·0.5LiMn0.42Ni0.42Co0.16O2 'composite' layered cathode material for lithium-ion batteries. *RSC Adv.* **2,** 8797 (2012).
28. Fell, C. R. *et al.* High pressure driven structural and electrochemical modifications in layered lithium transition metal intercalation oxides. *Energy Environ. Sci.* **5,** 6214 (2012).
29. Sun, L., Marrocchelli, D. & Yildiz, B. Edge dislocation slows down oxide ion diffusion in doped $CeO_2$ by segregation of charged defects. *Nat. Commun.* **6,** 6294 (2015).
30. Qiu, B. *et al.* Gas–solid interfacial modification of oxygen activity in layered oxide cathodes for lithium-ion batteries. *Nat. Commun.* **7,** 12108 (2016).

# Supplementary Information
**Supplementary Figures**

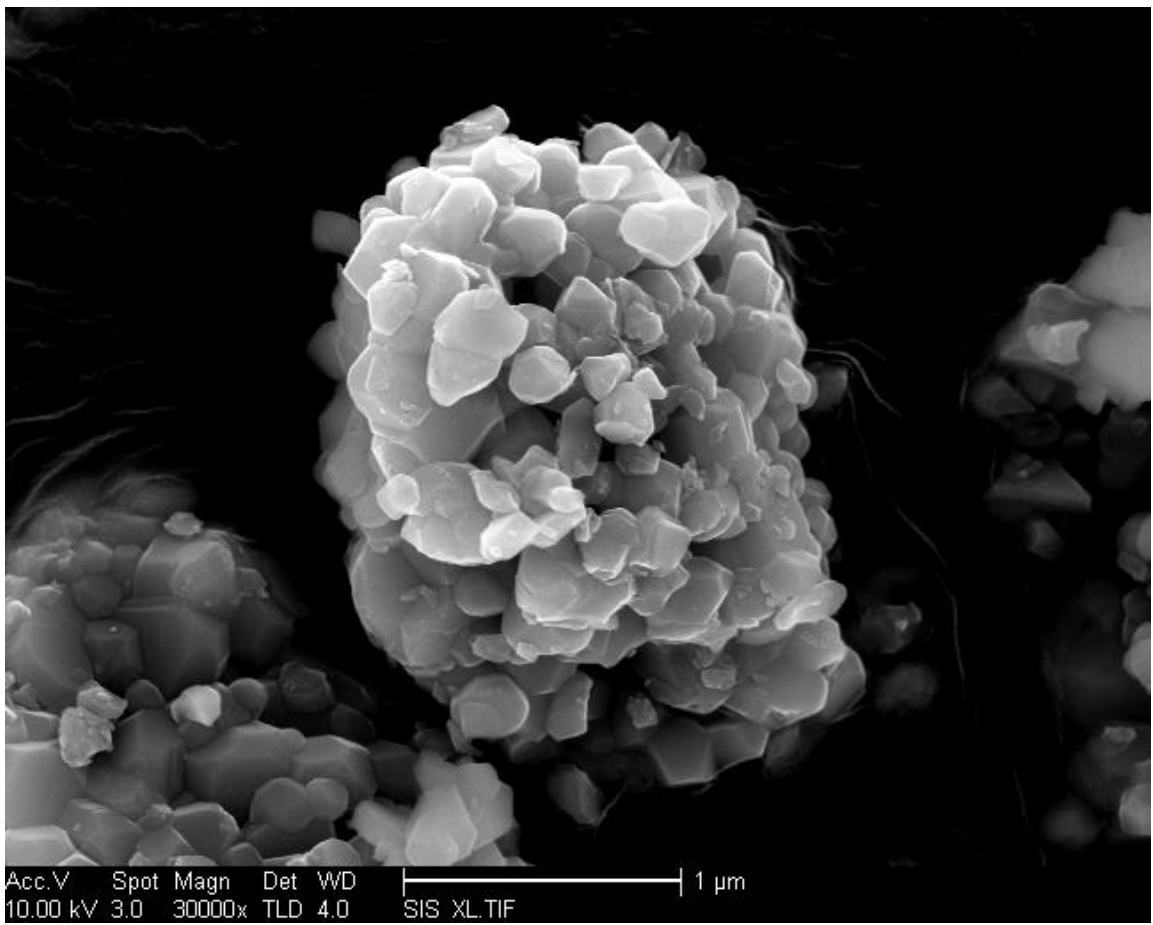

**Figure S1: Scanning electron microscopy image of the pristine LRLO material**. The primary particles (sub micron single crystallites) are agglomerated to larger secondary particles to improve the volumetric energy density.

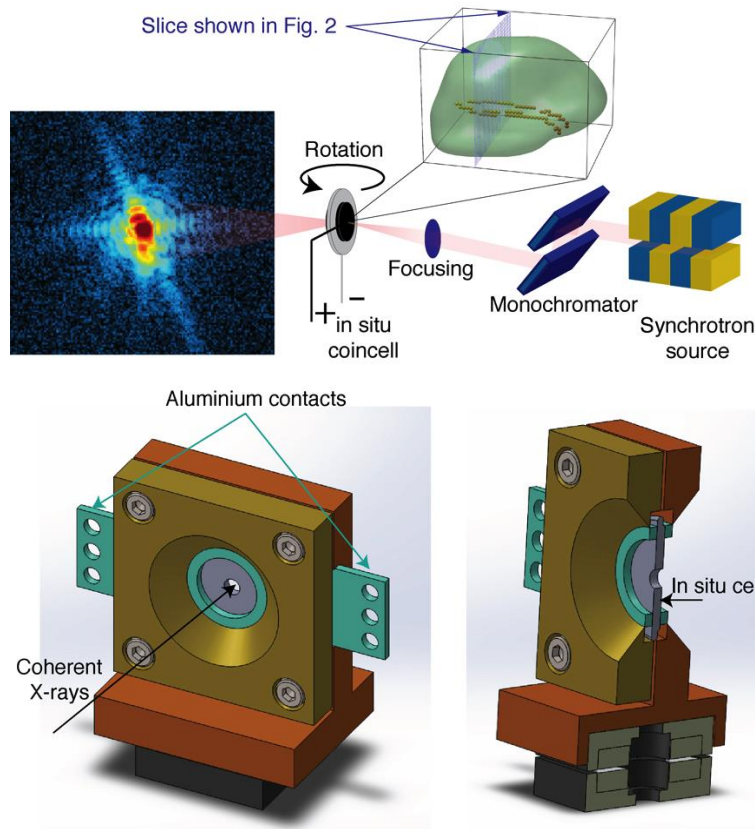

**Figure S2: Experimental description.** Schematic of the experimental arrangement (top, the slice through the 3D volume of the particle shown in Fig. 2 is indicated) and the sample holder with the *in situ* coin cell used in both experiments (bottom).

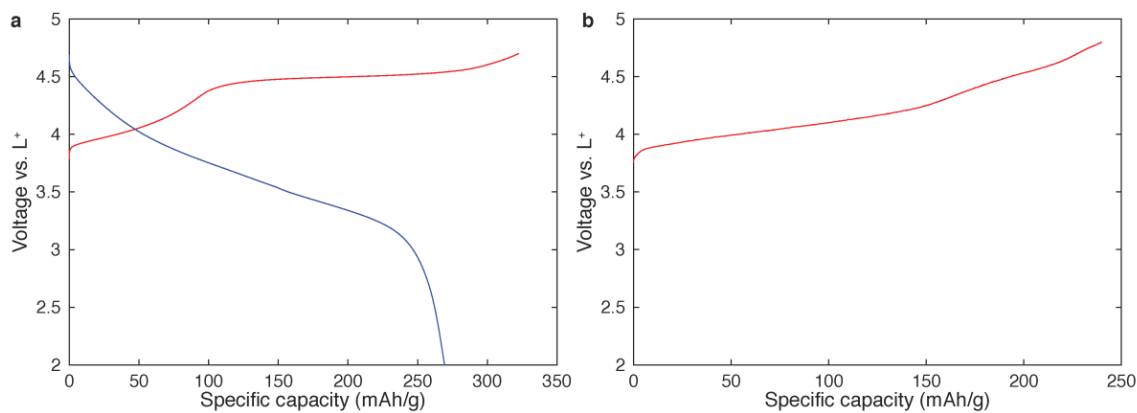

**Figure S3: Electrochemical performance. a** Specific capacity of the *in situ* LRLO cell measured simultaneously with the x-ray imaging data shown in Figs. 2 and S7. **b** Specific capacity of the *in situ* NCA cell measured simultaneously with the x-ray imaging data shown in Fig. 3.

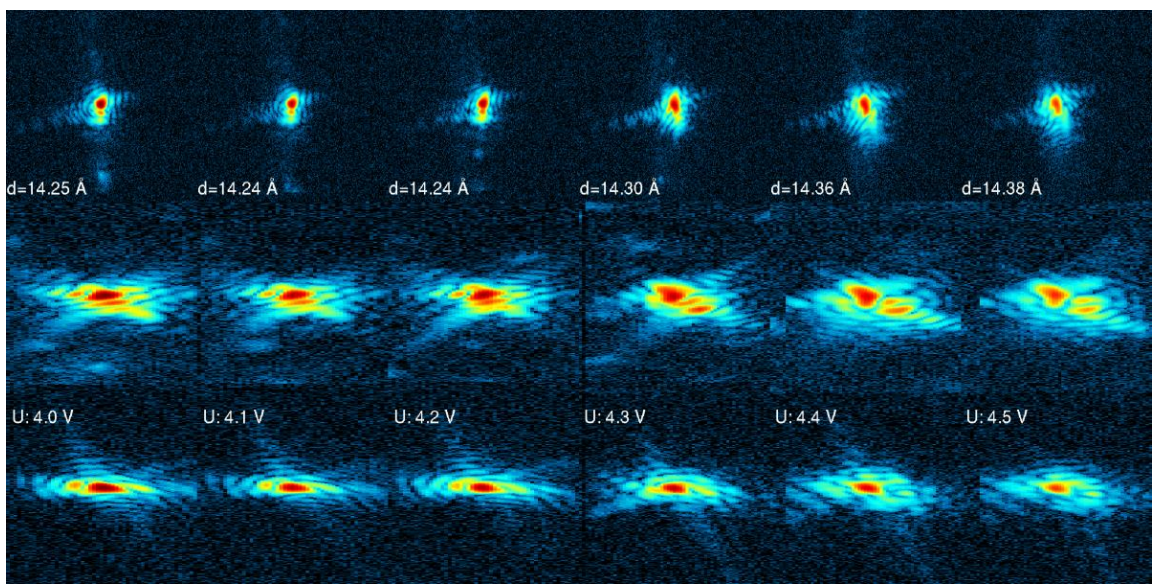

**Figure S4**: Diffraction data collected for the particle shown in Fig. 1 and 2 of the main text. The voltage and the average lattice parameter are indicated. The average lattice parameter was determined from the center of mass of the 3D reciprocal space data around the Bragg peak. The top row shows images as seen on the detector during the measurement, the whole pattern is 0.027 Å$^{-1}$ X 0.027 Å$^{-1}$ in size. The second row and third rows show two different orthogonal views of the 3D diffraction pattern collected during the rocking series. The scattering vector is approximately vertical (top), out of plane (middle), horizontal (bottom).

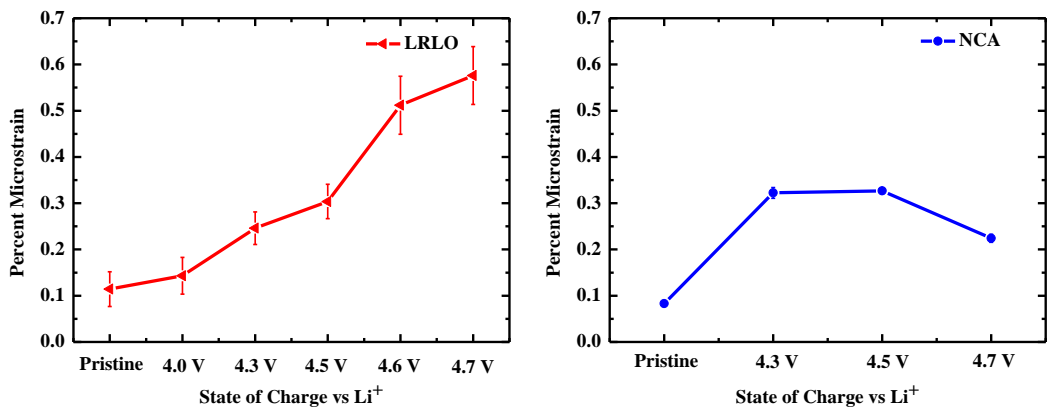

**Figure S5: Comparison of microstrain changes for different states of charge of LRLO and NCA samples determined from the Williamson-Hall analysis of the Bragg peak widths from conventional x-ray diffraction experiments on a large number of particles.** The data for LRLO (a) was measured *in situ*, while NCA (b) was measured *ex situ*.

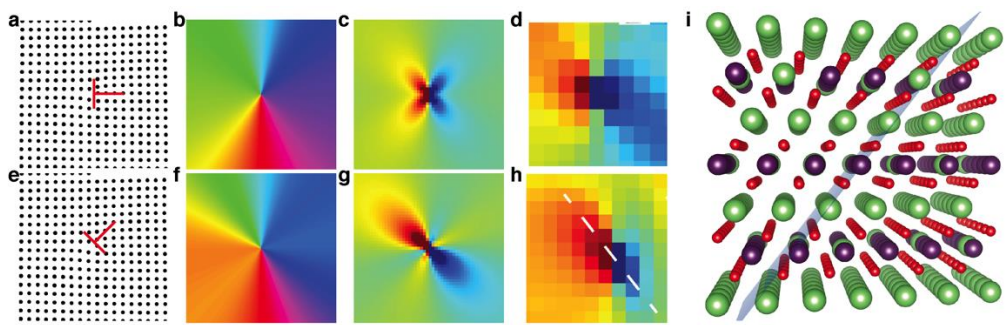

**Figure S6: Microscopic model of the displacement field and strain around a dislocation. a** Schematic of an edge dislocation with extra half plane parallel to the layers and the corresponding displacement field **b** and strain field **c**. **d** Strain field around a dislocation in NCA from Fig. 3 (4.2 V). **e** A schematic of an edge dislocation with the extra half plane building an angle with the layers and the corresponding displacement field **f** and strain **g**. **h** Strain field around a dislocation in LRLO from Fig. 2 (4.3 V). **i** The geometry of the possible orientation of the extra half plane {403} in LRLO is indicated by semitransparent blue plane through the center of the image.

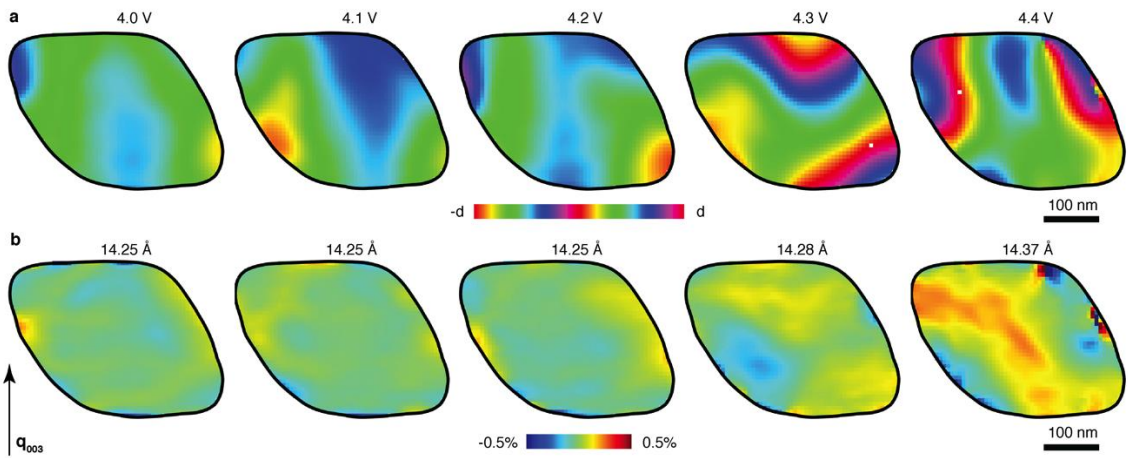

**Figure S7:** *In situ* **evolution of a second LRLO nanoparticle during charge. a** The changes in the displacement field along the (001) direction in a plane through the nanoparticle during charge. The voltage vs. Li⁺ is indicated at the top. Edge dislocations emerge at 4.4 V vs. Li⁺. A screw dislocation is present in this nanoparticle in the pristine state and through the charge. **b** The strain along the (001) direction inside of the nanoparticle calculated from the 3D displacement fields in **a**. The strain is shown around the average lattice constant indicated at the top.

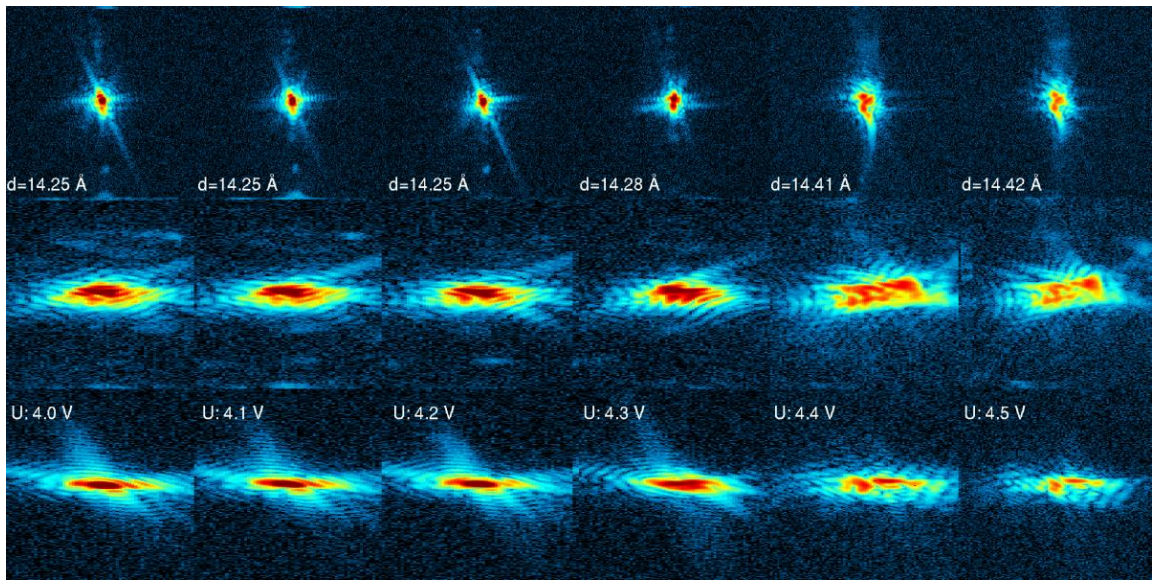

**Figure S8**: Diffraction data collected for the particle whose microstrain is shown in Fig. 4 in the main text by magenta symbols. The reconstructions are shown in Fig. S7. The voltage and the average lattice parameter determined from the center of the diffraction peak are indicated. The top row shows images as seen on the detector during the measurement, the whole pattern is 0.027 Å⁻¹ X 0.027 Å⁻¹ in size. The second row and third rows show two different orthogonal views of the 3D diffraction pattern collected during the rocking series. The scattering vector is approximately vertical (top), out of plane (middle), horizontal (bottom).

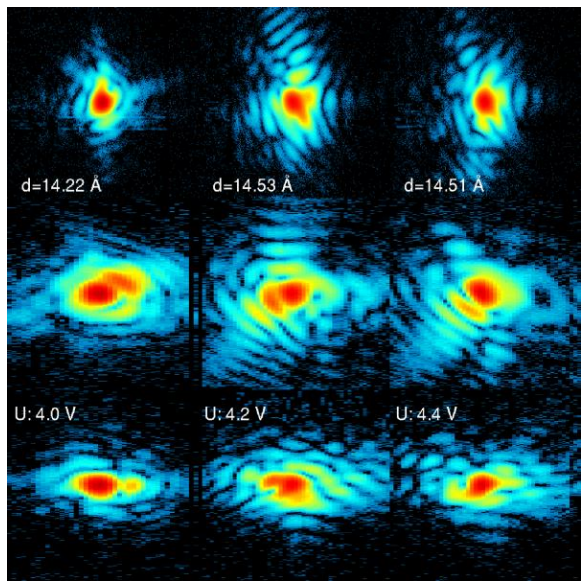

**Figure S9:** Diffraction data collected for the NCA particle shown in Fig. 3 in the main text. The top row shows images as seen on the detector, the whole pattern is 0.038 Å$^{-1}$ X 0.038 Å$^{-1}$ in size. The second row and third rows show two different orthogonal views of the 3D diffraction pattern collected during the rocking series. The scattering vector is approximately vertical (top), out of plane (middle), horizontal (bottom).

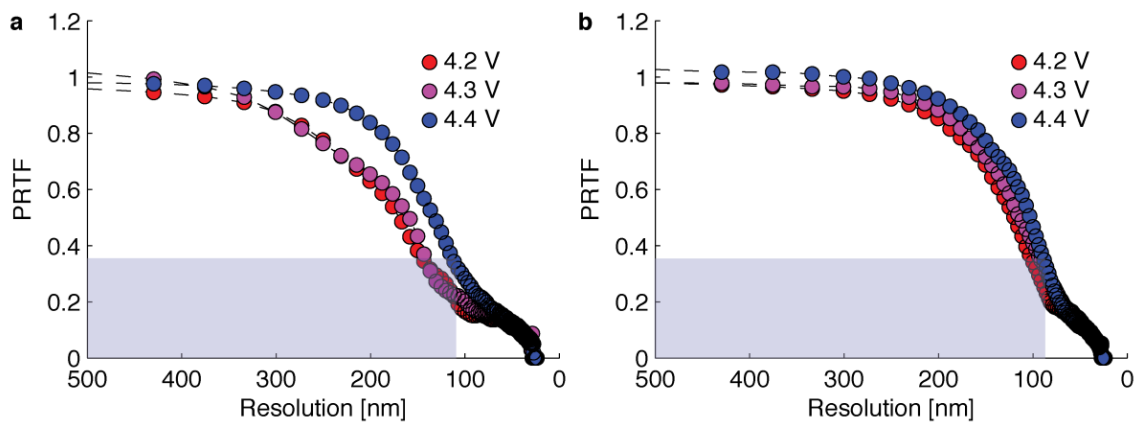

**Figure S10: Phase retrieval transfer function (PRTF). a** PRTF for the reconstructions shown in Fig. 2 of the main text. **b** PRTF for the particle shown in Fig. S7.

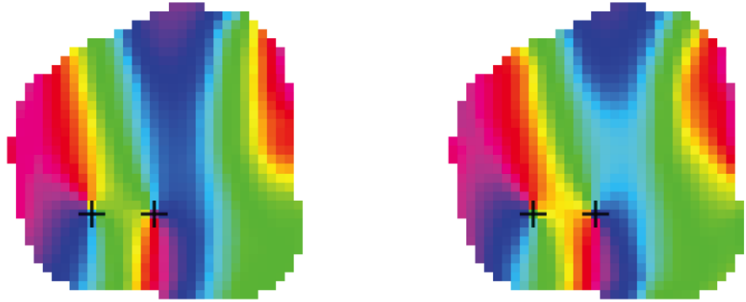

**Figure S11: Comparison between two different reconstructions of the same rocking scan.** Note that although the results from reconstruction procedures initiated from different random starts differ slightly, which is reflected in the PRTF resolution of about 100 nm, the positions of the dislocations are consistent between different reconstructions to a much better resolution (within a pixel, i.e. 7.75 nm).

**Experimental details**

The design of the *in situ* cells is described elsewhere [1,2]. The batteries were mounted on sample holders manufactured using a 3D printer (see Fig. S2). The electrode was prepared with 80 % of active material 10 % binder and 10 % carbon black. The cells were charged to a 4.1 V and discharged to 3.5 V three times, which improved the stability and electrochemical response of the nanoparticles. Charging to these low voltages is completely reversible. The experiments on lithium-rich layered oxides (LRLO) were conducted at 34 ID beamline of the Advanced Photon Source (Argonne National Laboratory, ANL, USA). A photon energy of 9 keV and a sample-to-detector distance of 60 cm were used in the experiments. The experiments on NCA were conducted at P10 beamline of PETRA III (Deutsches Elektronen Synchrotron, DESY, Germany) with a photon energy of 9.6 keV and a sample-to-detector distance of 182 cm. Timepix (34ID) or Lambda (P10) 2D detectors were used both with a pixel size of 55 μm x 55 μm. Figure S3 shows the electrochemical profiles measured simultaneously with the x-ray data on the in situ cell. Figures S4, S8, S9 show slices through typical x-ray diffraction profiles.

**X-ray data collection and reconstruction procedure**

In all experiments rocking scans, ~1° wide with 30-60 points were collected. During the analysis, various reconstruction procedures were tested, including combinations of ER alternating with HIO, RAAR, or DM [3,4]. The diffraction data was cropped (APS data) or binned (PETRA III data) to 128x128 pixels before running reconstructions. The number of iterations was varied between 410-2000. All attempts resulted in very similar reconstructions. In this work we used an average of 10 results, each being an average of 25 reconstructions retrieved in a guided procedure developed in [5] (5 generations, 50 population). The reconstructions were run using a GPU optimized code on multiple GeForce Titan Black graphics cards.

**Identifying the defect network**

The dislocations were identified using the method presented in [5]. The dislocation density was calculated by dividing the integrated dislocation length of the network by the volume of the nanoparticle. Although the resolution defined by the phase retrieval transfer function is about 100 nm, the positions of the dislocations are resolved to a much higher precision (see Figs. S10 and S11), possibly because the core of the dislocation is determined by the long-range displacement field.

**Model calculations within the isotropic elastic model**

The displacement and strain fields around dislocations provide unique access to the local elastic properties of the material as well as the exact nature of dislocations [7,8]. We have used the isotropic model to calculate the displacement field around an edge dislocation [7]

$$u_x = \frac{b}{2\pi}\left[\tan^{-1}\left(\frac{y}{x}\right) + \frac{xy}{2(1-\nu)(x^2+y^2)}\right]$$

$$u_y = -\frac{b}{2\pi}\left[\frac{1-2\nu}{4(1-nu)}\log(x^2+y^2) + \frac{x^2-y^2}{4(1-\nu)(x^2+y^2)}\right]$$

where $u_x$ and $u_y$ are the displacements perpendicular and parallel to the extra plane, respectively, b is the modulus of the Burgers vector, ν is the Poisson ratio (assumed 0.3 here [9]). The strain was calculated as the derivative of the displacement field along the q vector, $u_q$. For a Burgers vector parallel to q, $u_q=u_x$, and the resulting displacement and strain are shown in Fig. S6a-c, which is in reasonable agreement with the measurements in NCA (see Fig. S6d). The displacement and strain field in LRLO is not well reproduced with this model calculation (see Fig. S6h), and we identify the {403} planes (see Fig. S6i) as likely candidates to form the half plane dislocations, as in this plane oxygen vacancies are more likely to develop [10]. This plane has an angle of 52 ° with the layers. Fig. S6e-h shows the corresponding displacement field, which agrees qualitatively with the measurements in LRLO.

**Microstrain and superstructure analysis**

The miscrostrain was analyzed by examining line broadening observed in the synchrotron X-ray diffraction (SXRD) patterns, which were collected at the Advanced Photon Source at Argonne National Laboratory. The wavelength of the x-ray source was 0.11165 Å for LRLO and 0.414215 Å for NCA samples. The data for LRLO was measured *in situ*, while NCA was measured *ex situ*. Williamson−Hall analysis of all peaks that exhibit the best linear fitting was carried out for a quantification of microstrain changes during initial charging process [6]. The instrumental broadening was corrected based on the standard sample $CeO_2$ using the following equation:

$$FW(S)^D = FWHM^D - FW(I)^D, \qquad (1)$$

where *FWHM* is measured full width at half maximum of each peak; *FW*(s) and *FW*(I) are the calculated full-widths for the sample and instrument, respectively; D is the deconvolution parameter which is set to be 1.5 for all the analysis. The sample broadening is ascribed to particle size and microstrain based on the Williamson–Hall method by the following equation:

$$FW(S) \times \cos(\theta) = \frac{K \times \lambda}{Size} + 4 \times Strain \times \sin(\theta), \qquad (2)$$

where *K* is the crystallite shape factor and was assumed to be 0.9; $\theta$ is the diffraction angle; and $\lambda$ is the x-ray wavelength of the source. The microstrain is extracted from the slope of the plot of *FW*(s) cos($\theta$) versus 4 sin($\theta$). For LRLO sample, the microstrain gradually increases until 4.5 V and keeps increasing even more rapidly after 4.5 V (see Fig. S5a). The maximum microstrain for NCA sample occurs around 4.3 V and the microstrain reduces upon further charging (see Fig. S5b).

The superstructure peak was analyzed in the data used for the Williamson-Hall analysis. The superstructure peak height was determined as the peak intensity after background subtraction and normalization (see right inset in Fig. 4).

**References**

[1]  A. Singer, A. Ulvestad, H.-M. Cho, J. W. Kim, J. Maser, R. Harder, Y. S. Meng, and O. G. Shpyrko, Nano Lett. **14**, 5295 (2014).

[2]  A. Ulvestad, A. Singer, J. N. Clark, H. M. Cho, J. W. Kim, R. Harder, J. Maser, Y. S. Meng, and O. G. Shpyrko, Science (80-. ). **348**, 1344 (2015).

[3]  V. Elser, J. Opt. Soc. Am. A **20**, 40 (2003).

[4]  D. R. Luke, Inverse Probl. **21**, 37 (2005).

[5]  J. N. Clark, J. Ihli, A. S. Schenk, Y.-Y. Kim, A. N. Kulak, J. M. Campbell, G.

Nisbet, F. C. Meldrum, and I. K. Robinson, Nat. Mater. **14**, 780 (2015).

[6] G. K. Williamson and W. H. Hall, Acta Metall. **1**, 22 (1953).

[7] D. Hull and D. J. Bacon, in *Introd. to Dislocations* (Elsevier, 2011), p. iv.

[8] A. Ulvestad, A. Singer, J. N. Clark, H. M. Cho, J. W. Kim, R. Harder, J. Maser, Y. S. Meng, and O. G. Shpyrko, Under Rev. Sci. (2014).

[9] Y. Qi, L. G. Hector, C. James, and K. J. Kim, J. Electrochem. Soc. **161**, F3010 (2014).

[10] B. Qiu, M. Zhang, L. Wu, J. Wang, Y. Xia, D. Qian, H. Liu, S. Hy, Y. Chen, K. An, Y. Zhu, Z. Liu, and Y. S. Meng, Nat. Commun. **7**, 12108 (2016).